\documentclass[a4paper,11pt]{article}
\usepackage{pos}
\usepackage{float}

\title{Exact NNLO corrections vs K-factors in PDF fits}
\ShortTitle{Exact NNLO corrections vs K-factors in PDF fits}

\author*[a,b,c]{Tanishq Sharma}

\affiliation[a]{Dipartimento di Fisica, Universit\`a di Torino and INFN, Sezione di Torino,\\
Via P.~Giuria 1, 10125 Torino, Italy}

\affiliation[b]{Department of Physics and Astronomy, Vrije Universiteit Amsterdam,\\
1081 HV Amsterdam, The Netherlands}

\affiliation[c]{Nikhef Theory Group,\\
Science Park 105, 1098 XG Amsterdam, The Netherlands}

\emailAdd{tanishq.sharma@unito.it}

\abstract{Parton distribution functions (PDFs) often include datasets corresponding to processes whereby the theoretical predictions at next-to-next-to-leading order (NNLO) in peturbative QCD have to be approximated,
and this approximation may be performed using K-factors, which in turn depend on the PDF set used to compute them. In this study, we investigate
the impact of K-factors produced with various PDF sets, namely CT18, MSHT20 and NNPDF4.0 on (differential) cross sections of top pair production at the Large Hadron Collider (LHC). Furthermore, we perform a new fit (otherwise analogous to NNPDF4.0 with MHOUs)
where the exact NNLO corrections are used in the fitting procedure and compare the K-factors obtained from this fit with those obtained from the above mentioned PDF sets. We find
good agreement amongst K-factors obtained from these different PDF sets.}

\FullConference{31st International Workshop on Deep Inelastic Scattering (DIS2024)\\
 8–12 April 2024\\
Grenoble, France\\}

\begin{document}
\maketitle

Higher statistics and a finer control over the sources of systematic uncertainties has led to a significant 
increase in the accuracy and precision of experimental data coming out of the LHC. This often leads to the
PDF uncertainty being one of the dominant sources of uncertainties in the theoretical predictions, which, together with the experimental data
allow for extraction of Standard Model (SM) parameters. This 
motivates a more accurate and precise determination of the PDFs. In particular, we aim to achieve a PDF accuracy of less than a percent. Percent level effects in the determination of PDFs can originate from
various theoretical sources. Consequently, improvements in PDFs can be achieved through various
means, including, but not limited to, the inclusion of Missing Higher Order Uncertainties (MHOUs)~\cite{NNPDF:2024dpb}, or moving towards approximate
next-to-next-to-next-to-leading order (aN$^3$LO) PDFs~\cite{NNPDF:2024nan, McGowan:2022nag}, both of which may improve accuracy, or by inclusion of pure NNLO corrections to processes
in favor of K-factor approximations, which may improve both, accuracy and precision. A potential drawback of the K-factor approximation approach may lie in the fact that K-factors are a blanket correction towards the next
perturbative order, without taking into account how different partonic channels' contributions vary at higher perturbative orders. There is no a priori reason to assume that all the partonic channels will 
have similar NNLO corrections as their NLO counterparts. Therefore, in this study, we focus on the inclusion of pure NNLO corrections in lieu of K-factors specifically in the case of top pair 
production at the LHC.

It has been possible to conduct this study due to MATRIX~\cite{Grazzini:2017mhc, Catani:2019iny, Catani:2019hip, Denner:2016kdg, 
Buccioni:2019sur, Buccioni:2017yxi, Barnreuther:2013qvf, Catani:2007vq} interfaced with PineAPPL~\cite{christopher_schwan_2024_12795745, Carrazza:2020gss}, which allows for the generation of
theoretical predictions for the hard partonic cross section of top pair production at NNLO, in the form of PineAPPL interpolation grids. The grids computed use a dynamic scale $H_T$ to maximize perturbative convergence (as
suggested in~\cite{Czakon:2016olj}). The use of interpolation grids allows for fast and efficient computation of theoretical predictions with any set of PDFs and at any 
perturbative order, thus allowing for a quick computation of K-factors for a given PDF set. The expression for the computation of 
a K-factor is given in Eq.~\ref{eq:kfactor}. 
\begin{equation}
\label{eq:kfactor}
k = \frac{\hat{\sigma}_{\text{NNLO}} \otimes PDF_{\text{NNLO}}}{\hat{\sigma}_{\text{NLO}} \otimes PDF_{\text{NNLO}}}
\end{equation}

In table~\ref{tab:datasets}, the top pair production datasets for which we perform the comparisons are listed. All of these datasets were included
in NNPDF4.0~\cite{NNPDF:2021njg} and NNPDF4.0 with MHOUs~\cite{NNPDF:2024dpb}.

\begin{table}[]
    \centering
    \small
    \begin{tabular}{ p{3.7cm} | c c c }
    \hline
    Dataset & Observable & N$_{dat}$ & Ref. \\
    \hline
    ATLAS $t\bar{t}$ 7 TeV & $\sigma_{t\bar{t}}$ & 1 & \cite{ATLAS:2014nxi} \\
    ATLAS $t\bar{t}$ 8 TeV & $\sigma_{t\bar{t}}$ & 1 & \cite{ATLAS:2014nxi} \\
    ATLAS $t\bar{t}$ 13 TeV & $\sigma_{t\bar{t}}$ & 1 & \cite{ATLAS:2020aln} \\
    ATLAS $t\bar{t}$ $2\ell$ 8 TeV & $1/\sigma\ d\sigma/d|y_{t\bar{t}}|$ & 5 & \cite{ATLAS:2016pal} \\ 
    ATLAS $t\bar{t}$ $\ell+$jets 8 TeV & $1/\sigma\ d\sigma/d|y_{t}|$ & 5 & \cite{ATLAS:2015lsn} \\
    ATLAS $t\bar{t}$ $\ell+$jets 8 TeV & $1/\sigma\ d\sigma/d|y_{t\bar{t}}|$ & 5 & \cite{ATLAS:2015lsn} \\
    CMS $t\bar{t}$ 5 TeV & $\sigma_{t\bar{t}}$ & 1 & \cite{CMS:2017zpm} \\
    CMS $t\bar{t}$ 7 TeV & $\sigma_{t\bar{t}}$ & 1 & \cite{CMS:2016yys} \\
    CMS $t\bar{t}$ 8 TeV & $\sigma_{t\bar{t}}$ & 1 & \cite{CMS:2016yys} \\
    CMS $t\bar{t}$ 13 TeV & $\sigma_{t\bar{t}}$ & 1 & \cite{CMS:2015yky} \\
    CMS $t\bar{t}$ $2\ell$ 8 TeV & $1/\sigma\ d^2\sigma/dm_{t\bar{t}}d|y_{t}|$ & 16 & \cite{CMS:2017iqf} \\
    CMS $t\bar{t}$ $\ell+$jets 8 TeV & $1/\sigma\ d\sigma/dy_{t\bar{t}}$ & 10 & \cite{CMS:2015rld} \\
    CMS $t\bar{t}$ $2\ell$ 13 TeV & $d\sigma/dy_{t}$ & 10 & \cite{CMS:2018adi} \\

    \hline
    \end{tabular}
    
    \caption{The top pair production experimental datasets from LHC for which we perform the comparisons.}
    \label{tab:datasets}
    \end{table}

\begin{figure}[]
    \centering
    \includegraphics[width=0.93\textwidth]{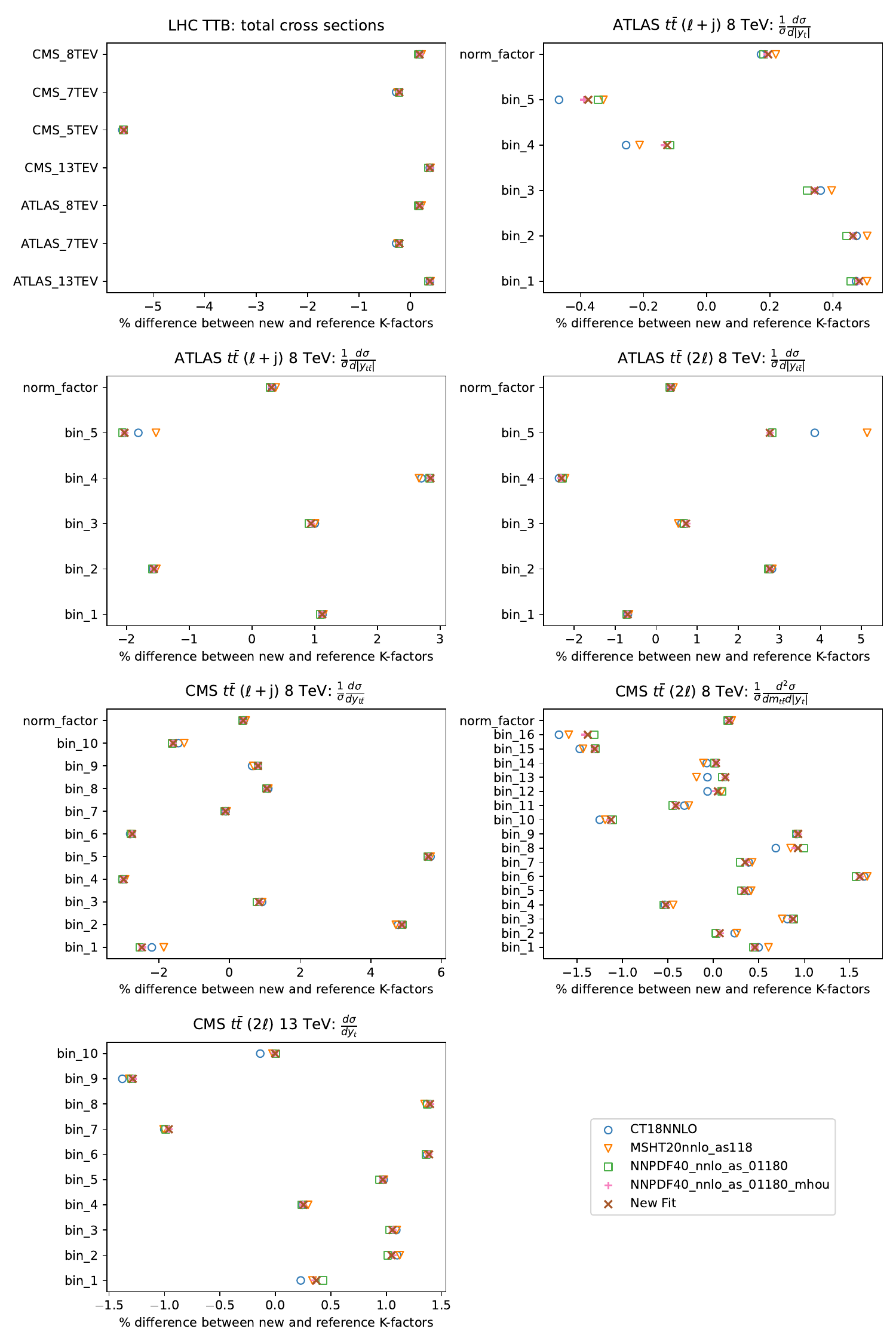}
    \caption{This figure shows the percentage difference between K-factors that were used in NNPDF4.0 (the reference K-factors)
and K-factors obtained from pure NNLO grids convolved with CT18~\cite{Hou:2019efy}, MSHT20~\cite{Bailey:2020ooq},
NNPDF4.0~\cite{NNPDF:2021njg}, NNPDF4.0 with MHOUs~\cite{NNPDF:2024dpb} and a New Fit
(which is otherwise analogous to NNPDF4.0 with MHOUs) where the exact NNLO $t\bar{t}$ grids are used during the fitting procedure.}
    \label{fig:plots}
\end{figure}


To perform a systematic comparison, we start with K-factors that were used in the determination of NNPDF4.0~\cite{NNPDF:2021njg} (and NNPDF4.0 with MHOUs~\cite{NNPDF:2024dpb}),
which were obtained using NLO predictions from \texttt{mg5\_aMC}~\cite{Alwall:2014hca}, and NNLO predictions from publicly available fastNLO tables~\cite{Czakon:2017dip, Czakon:2019yrx} and \texttt{top++}~\cite{Czakon:2011xx}.
These act as our reference K-factors. We proceed by taking the $t\bar{t}$ PineAPPL grids and convolving them with some select PDF sets, namely CT18~\cite{Hou:2019efy}, MSHT20~\cite{Bailey:2020ooq}, NNPDF4.0~\cite{NNPDF:2021njg}, NNPDF4.0 with MHOUs~\cite{NNPDF:2024dpb}.
Performing this convolution at NNLO and NLO allows for the computation of K-factors (using eq.~\ref{eq:kfactor}), where each set of K-factors depends on the specific PDF set used to compute them. In addition, we perform a new fit where the exact NNLO $t\bar{t}$ grids are used during the fitting procedure (and all else remains
same as NNPDF4.0 with MHOUs). We also compute K-factors for this new fit. In Fig.~\ref{fig:plots}, the percentage difference between the reference K-factors and the K-factors obtained
during this study are shown. The results vary from dataset to dataset with the percentage difference w.r.t. the reference K-factors, going as high as 5-6\% for one data point. However, the percentage differences between the 
K-factors computed using the select PDFs (and the new fit) are extremely small, indicating a consistency between the K-factors obtained using different PDF sets. This demonstrates that for the large part,
K-factors are able to capture the NNLO corrections fairly well, and as we move towards the use of exact NNLO corrections for top pair production in the fitting procedure, it is reasonable to expect the impact
on the PDFs to be minimal.

\bibliographystyle{apsrev}
\bibliography{cite}

\end{document}